\documentclass[aps,prb,superscriptaddress,reprint,amsmath,amssymb]{revtex4-1}

\usepackage{graphics}
\usepackage[pdftex]{graphicx}
\usepackage{tabularx}
\graphicspath{{./Figures/}}
\usepackage{longtable}
\usepackage{amsmath}
\usepackage{color,soul}
\usepackage{braket}
\usepackage{float} 

\begin{document}

\title{Increasing the performance of the superconducting spin valve using a Heusler alloy}

\author{A.A.~Kamashev}
\affiliation{Zavoisky Physical-Technical Institute, Russian Academy of Sciences, 420029 Kazan, Russia}
\affiliation{Leibniz Institute for Solid State and Materials Research IFW Dresden, D-01171 Dresden, Germany}

\author{A.A.~Validov}
\affiliation{Zavoisky Physical-Technical Institute, Russian Academy of Sciences, 420029 Kazan, Russia}

\author{J.~Schumann}
\affiliation{Leibniz Institute for Solid State and Materials Research IFW Dresden, D-01171 Dresden, Germany}

\author{V.~Kataev}
\affiliation{Leibniz Institute for Solid State and Materials Research IFW Dresden, D-01171 Dresden, Germany}

\author{B.~B\"{u}chner}
\affiliation{Leibniz Institute for Solid State and Materials Research IFW Dresden, D-01171 Dresden, Germany}
\affiliation{Institute for Solid State Physics, Technical University Dresden, D-01062 Dresden, Germany}

\author{Y.V.~Fominov}
\affiliation{L.~D.\ Landau Institute for Theoretical Physics, Russian Academy of Sciences, 142432 Chernogolovka, Russia}
\affiliation{National Research University Higher School of Economics, 101000 Moscow, Russia}

\author{I.A.~Garifullin}
%\email{ilgiz0garifullin@gmail.com}
\affiliation{Zavoisky Physical-Technical Institute, Russian Academy of Sciences, 420029 Kazan, Russia}

%\date{\today}

\begin{abstract}
We have studied superconducting properties of the spin-valve thin
layer heterostructures CoO$_x$/F1/Cu/F2/Cu/Pb where the
ferromagnetic F1 layer was standardly made of Permalloy whereas for
the F2 layer we have taken a specially prepared film of the Heusler
alloy Co$_2$Cr$_{1-x}$Fe$_x$Al with a small degree of spin
polarization of the conduction band. The heterostructures
demonstrate a significant superconducting spin-valve effect, i.e. a
complete switching on and off of the superconducting current flowing
through the system by manipulating the mutual orientations of the
magnetization of the F1 and F2 layers. The magnitude of the effect
is doubled in comparison with the previously studied analogous
multilayers with the F2 layer made of the strong ferromagnet Fe.
Theoretical analysis shows that a drastic enhancement of the
switching effect is due to a smaller exchange field in the
heterostructure coming from the Heusler film as compared to Fe. This
enables to approach almost ideal theoretical magnitude of the
switching in the Heusler-based multilayer with the F2 layer
thickness of $\sim 1$\,nm.
\end{abstract}

%\pacs{74.45+c, 74.25.Nf, 74.78.Fk}


\maketitle

\section{Introduction}

Historically, the first idea manipulate the $T_c$ of a superconductor by sandwiching it between two ferromagnetic
insulators is due to de Gennes \cite{deGennes1966}. Regarding the case of metallic ferromagnets the physical
principle of the superconducting spin valve (SSV) is based on the idea proposed
by Oh {\it et al.} in 1997 \cite{Oh1997} who calculated the pairing wave function amplitude
in a trilayer F1/F2/S (where F1 and F2 are ferromagnetic layers and S is a superconducting layer)
and found out that the superconducting (SC) transition temperature $T_c$ depends on the mutual
orientation of the magnetizations $\boldsymbol{M_{\rm 1}}$ and $\boldsymbol{M_{\rm 2}}$ of the F1 and F2 layers.
Later, another construction based on three-layer thin films F1/S/F2 was proposed also theoretically \cite{Tagirov,Buzdin}.
According to the above theories,
for the parallel (P) configuration of $\boldsymbol{M_{\rm 1}}$ and $\boldsymbol{M_{\rm 2}}$ the transition
temperature $T_c^\mathrm{P}$ should be always smaller than $T_c^\mathrm{AP}$ for the antiparallel (AP)
orientation of the magnetic vectors because in the former case the mean exchange field from
the F-layers destructively acting on the Cooper pairs is larger. Thus, under favorable conditions
the switching between AP and P configurations, which could be achieved by an appropriate application
of a small external magnetic field, should yield a complete switching on and off of the superconducting current in such a construction.
	
A number of experimental studies have confirmed the predicted effect of the mutual orientation
of magnetizations in the F1/S/F2 structure on $T_c$ (see, e.g., \cite{Gu,Potenza,Moraru,Moraru2,Miao}).
However, the major difficulty in a practical realization of an SSV -- to get the difference
between $T_c$ for the AP and P geometries $\Delta T_c=T_c^\mathrm{AP}-T_c^\mathrm{P}$ larger
than the width $\delta T_c$ of the superconducting transition for a given configuration
of $\boldsymbol{M_{\rm 1}}$ and $\boldsymbol{M_{\rm 2}}$ -- was not overcome in these works.
One should note that the reported [Fe/V]$_n$ antiferromagnetically coupled superlattice \cite{Westerholt}
in which  $\Delta T_c$ could implicitly reach up to 200\,mK  cannot be considered as an SSV
because this system can not be switched from the AP to P orientation of the magnetizations instantaneously.

In addition to that, the effect of SSV becomes more complicated due to the following fact~\cite{Fominov2010}.
It is well known (see, for example, the review \cite{Bergeret2005}) that in the ferromagnetic layer the
Cooper pair acquires a nonzero momentum due to the Zeeman splitting of electronic levels. Its wave function
oscillates in the space when moving away from the S/F interface. If the F layer is thin enough, the wave
function is reflected from the surface opposite to the S/F interface. The interference of the incident and
reflected functions arises. Depending on the thickness of the F layer, the interference at the S/F interface
can be constructive or destructive. This should lead to an increase or decrease of the $T_c $ of the S/F
structure depending on the interference type.
	
From the experimental point of view, the results obtained for both theoretical designs of the SSV suggested
that the scheme by Oh {\it et al.} \cite{Oh1997} may be the most promising for the realization of the full SSV effect.
Indeed, this approach turns out to be successful. Previously we
have demonstrated a full switching between the normal and
superconducting states for the CoO$_x$/Fe1/Cu/Fe2/In spin-valve
structure \cite{Leksin2010}.  Later on we replaced the
superconducting In by Pb in order to improve superconducting
parameters \cite{Leksin2012a} and introduced an
additional technical Cu interlayer (N2) in order to prevent
degradation of the samples \cite{Leksin2013}. Thus, the final design
of the SSV structures was set as AFM/F1/N1/F2/N2/S. In this
construction the Cu interlayer (N1) decouples magnetizations of the
Fe1 (F1) and Fe2 (F2) layers and the antiferromagnetic (AFM) CoO$_x$
layer biases the magnetization of the Fe1 layer by anisotropy
fields. Despite substantial experimental efforts in optimizing the
properties of the In- and Pb-based SSVs
\cite{Leksin2012,Leksin2015}, in particular in reducing the width
$\delta T_c$,  our theoretical analysis of the properties of such
multilayers in the framework of the theory of Ref.~\cite{Fominov2010} has shown that the experimentally achieved
magnitude $\Delta T_c$ of the SSV effect  of 20\,mK and 40\,mK for
the two types of the S layer, respectively, was substantially
smaller as expected on theoretical grounds. Recently the interest to
the SSV increased considerably (see, e.g., review \cite{Linder2015}
and very recent publications
\cite{Flokstra2015,Alidoust,Mironov2015,Flokstra2016,Halterman2016,Srivastava})

Here, we present experimental results that evidence  a
significant improvement of the magnitude of the switching $\Delta
T_c$ in the Pb-based SSV by using the ferromagnetic Heusler alloy
(HA) Co$_2$Cr$_{1-x}$Fe$_x$Al as a material for the F2 layer.
Prepared under special well-defined conditions \cite{Kamashev2017}
the HA layer produces a substantially smaller exchange field acting
on the superconducting Cooper pairs as compared to the Fe layer of
the same thickness. This opens a possibility to grow
heterostructures where the theoretically desired parameters for the
maximum SSV effect could be practically realized yielding the
doubling of the magnitude of the SSV effect up to the almost ideal
theoretical value.

\section{Results}

\begin{figure}[t]
\center{\includegraphics[width=0.7\linewidth]{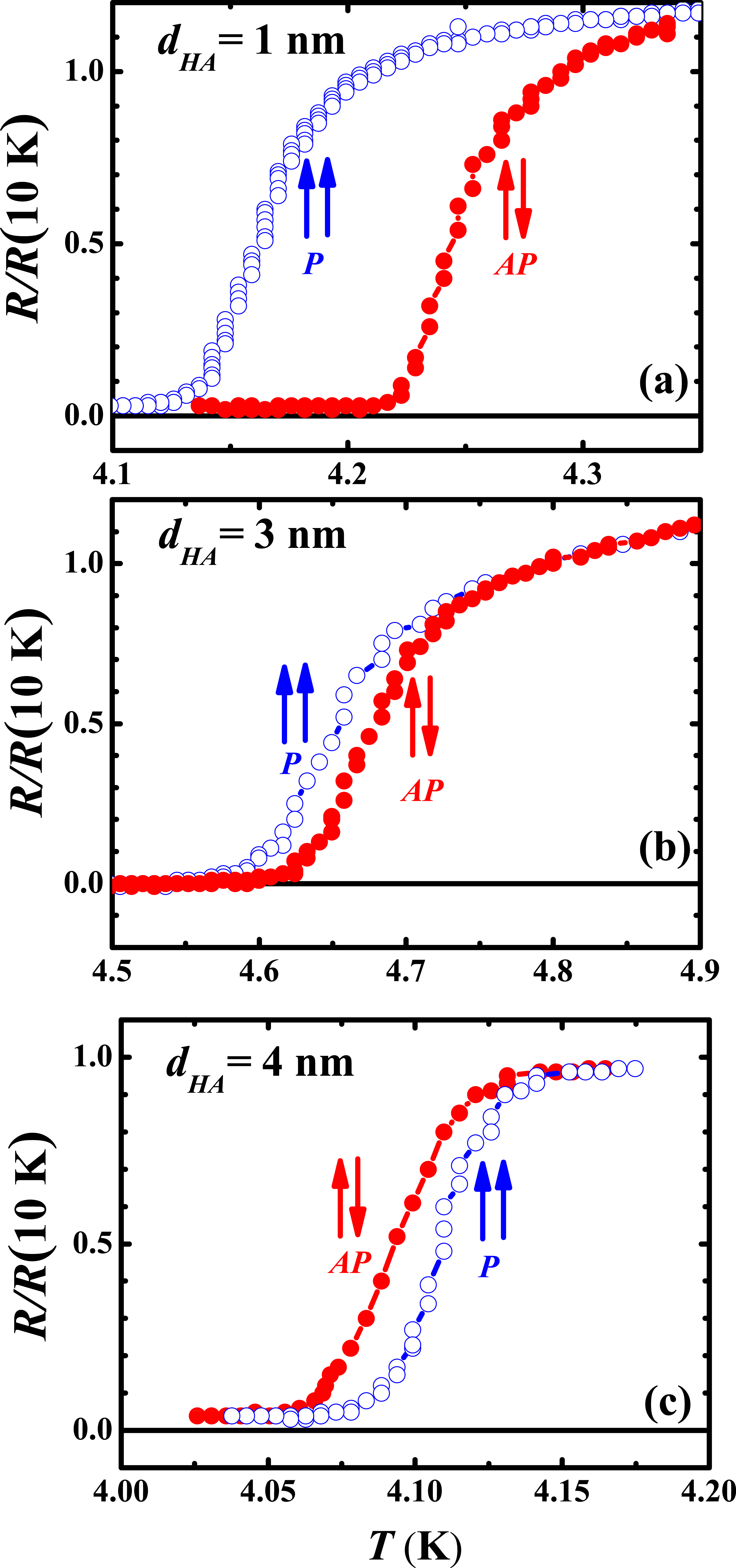}}
\caption{(Color online) Superconducting transition curves for
CoO$_x$/Py(5)/\-Cu(4)/Co$_2$Cr$_{1-x}$Fe$_x$Al/ Cu(1.5)/Pb(50)
multilayers with different thickness of the HA layer $d_{\rm HA}$
for P (open circles) and AP (closed circles) mutual orientation of
the  magnetizations $\boldsymbol{M_{\rm 1}}$ and $\boldsymbol{M_{\rm
2}}$ of the Py and Co$_2$Cr$_{1-x}$Fe$_x$Al$_x$ ferromagnetic
layers, respectively: (a) $d_{HA}=1$\,nm; (b) $d_{HA}=3$\,nm; (c)
$d_{HA}=4$\,nm.} \label{SCcurves}
\end{figure}

Technical particularities of the fabrication of the  SSV heterostructures which
have been studied in the present work have been reported in detail previously
(see Supporting Information). The new HA-based part of the multilayer
F2/N2/S = HA/Cu/Pb has been investigated in detail with the focus on the S/F
proximity effect very recently. It was shown \cite{Kamashev2017} that the degree
of the spin polarization of the conduction band of the HA film amounts to 30 \% for
the films prepared at a particular substrate temperature of $T_{\rm sub}$=300 K
during the growth of the HA layer and  to 70 \% at $T_{sub}$=600 K. In the AFM/F1/N1/F2/N2/S
structure it would be advantageous to achieve the penetration depth of the
Cooper pairs into F2 ferromagnetic layer should be as large as possible.
This means that the spin polarization of the conduction band should be small.
To fulfill this requirement we have prepared a set of samples
CoO$_x$/Py(5\,nm)/\-Cu(4\,nm)/Co$_2$Cr$_{1-x}$Fe$_x$Al/ Cu(1.5\,nm)/Pb(50\,nm)
with the HA layer of different thickness grown at $T_{\rm sub}= 300$\,K.
Representative superconducting transition curves are shown in Fig.~\ref{SCcurves}.

\begin{figure}[h]
\center{\includegraphics[width=1\linewidth]{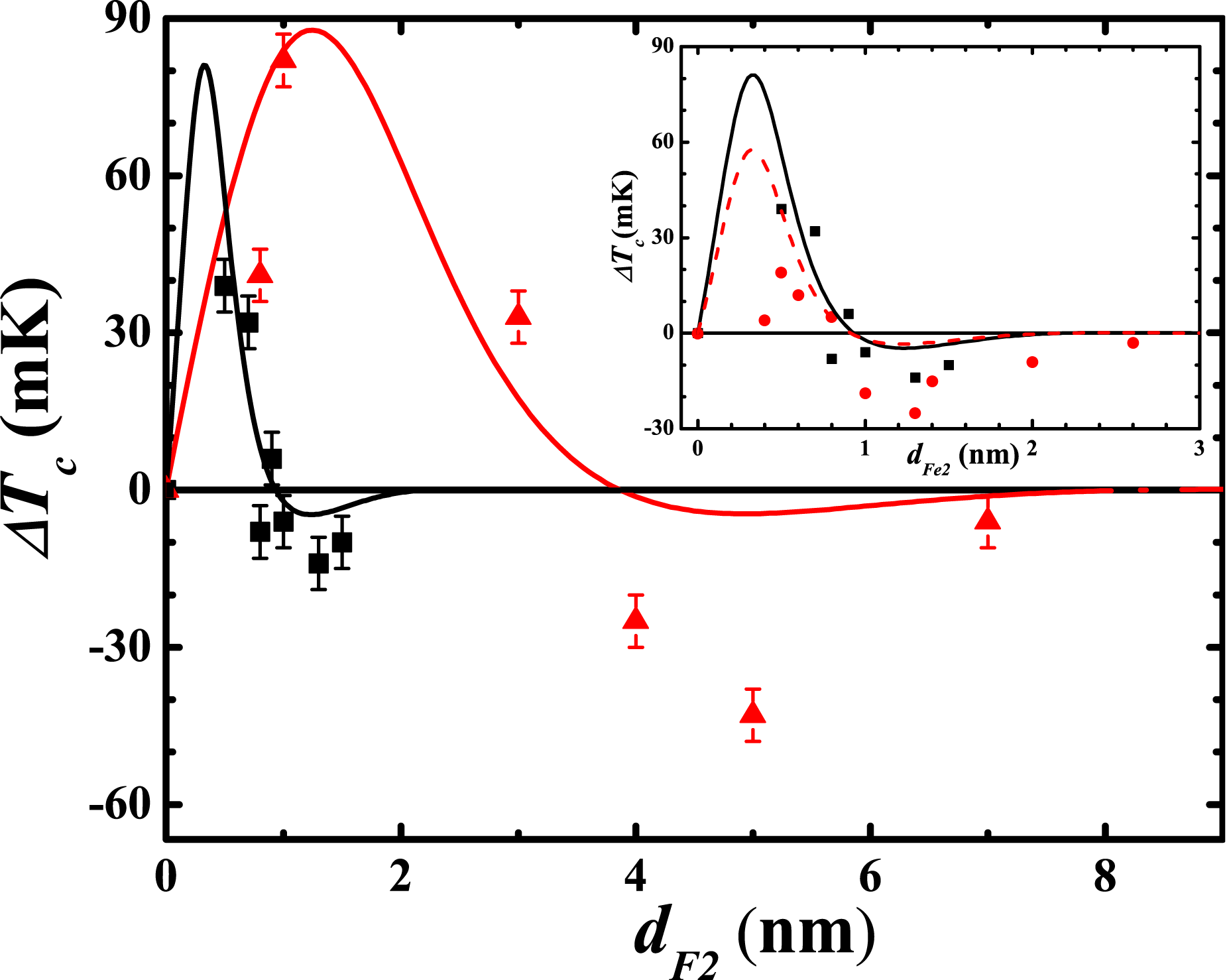}}
\caption{(Color online) Dependence of the $T_c$ shift $\Delta T_c =
T_c^\mathrm{AP} - T_c^\mathrm{P}$ on the F2-layer thickness $d_{\rm F2}$ in the
SSV heterostructures AFM/F1/N1/F2/N2/S. Triangles are the data
points for
CoO$_x$/Py(5)/\-Cu(4)/Co$_2$Cr$_{1-x}$Fe$_x$Al$_x$/ Cu(1.5)Pb(50)
from the present work. For comparison previous results for
CoO$_x$/Fe1/Cu/Fe2/Cu/Pb multilayers \cite{Leksin2015}  are plotted
with squares in the main panel and in the insert where additionally
also the data for the CoO$_x$/Fe1/Cu/Fe2/In SSV from
Ref.~\cite{Leksin2012} are plotted with circles for comparison.
Solid and dashed lines present the results of theoretical
modelling. (see the text)}
\label{HAdepend}
\end{figure}

A clear shift of the curves upon switching  the mutual orientation
of the magnetizations $\boldsymbol{M_{\rm 1}}$ and
$\boldsymbol{M_{\rm 2}}$ of the ferromagnetic layers between P and
AP configurations characteristic of the SSV effect is clearly
visible. The superconducting transition temperature was determined
as a midpoint of the transition curve. The dependence of the magnitude $\Delta T_c$ of the SSV
effect on the thickness of the HA layer is presented in
Fig.~\ref{HAdepend}.
The dependence $\Delta T_c(d_{\rm HA})$ reveals  an oscillating
behavior due to the interference of the Cooper pair wave functions
reflected from both surfaces of the ferromagnetic F2 layer (of the
order of 4 nm) proximate to the superconducting layer. This yields
for certain thicknesses of the F2 layer an inverse SSV effect
$\Delta T_c < 0$ \cite{Leksin2011}. The most remarkable result of
the present study is the magnitude of the direct SSV effect which
reaches for $d_{\rm HA}=1$\,nm (about two monolayers of HA) the maximum value of 80\,mK
(triangles in Fig.~\ref{HAdepend}) which surpasses the result for the analogous
heterostructure with Fe as the F2 layer \cite{Leksin2015} by a
factor of 2 (see the data comparison in Fig.~\ref{HAdepend}). As we
will discuss below, the achieved SSV effect in the Pb-based
heterostructure with the HA layer approaches the maximum value
predicted by theory. The scattering of $\Delta T_c$ is mainly due to some uncertainty
in the determination of the thickness of the HA layer which indirectly affects the
accuracy of the determination of $\Delta T_c$.

\section{Discussion}

To set up the basis for discussion we fist summarize the parameters
of the theory \cite{Fominov2010} describing
the SSV effect in the above systems. As described in
Ref.~\cite{Leksin2015}, in order to estimate these parameters
characterizing the properties of the S layer we use our experimental
data on the resistivity and the dependence of the $T_c$ on the S
layer thickness at a large unchanged thickness of the F layer in the
S/F bilayer. The residual resistivity $\rho_S =\rho(T_c)$ can be
determined from the residual resistivity ratio
$RRR$\,=\,R($T\,=\,300$\,K)/R($T_c$\,K)=[$\rho$(300\,K)+$\rho(T_c)$]/$\rho(T_c)$.
Since the room-temperature resistivity of the Pb layer is dominated
by the phonon contribution $\rho_{ph}$(300 K)=21 $\mu \Omega \cdot$cm
(see, e.g., \cite{Kittel}) we obtain $\rho_S$ values presented in Table~\ref{tc_tab}.
Then with the aid of the Pippard relations \cite{Pippard} the
following equality can be obtained \cite{Leksin2015}:
\begin{equation}
\label{eq1}
\rho_S l=\left( \frac{\pi k_B}{e} \right)^2 \frac{1}{\gamma^e v_F}.
\end{equation}
Here $\gamma^e$ denotes the electronic specific heat coefficient,
$v_F$ is the Fermi velocity of conduction electrons, and $l$ is the
mean-free path of conduction electrons. Using for Pb
$\gamma^e =1.6 \times 10^3$ erg/K$^2$cm$^3$ (Ref,~\cite{Kittel})
from Eq.~\ref{eq1} one can find the
mean-free path $l_S$, the diffusion coefficient of conduction
electrons $D_S$ and the superconducting coherence length
\begin{equation}
\label{eq2}
\xi_S=\sqrt{{\hbar D_S}\over {2\pi k_B T_{cS}}}.
\end{equation}
For the F layers the same procedure can be applied taking  into
account the definition of the superconducting coherence length in the
F layer \cite{Fominov2002}
\begin{equation}
\label{eq3}
\xi_F=\sqrt{{\hbar D_F}\over {2\pi k_B T_{cS}}},
\end{equation}
where $D_F$ is the diffusion coefficient for conduction  electrons
in the F layer and $T_{cS}$ is the superconducting transition temperature for an
isolated S layer.

The theory contains also the material-specific parameter  $\gamma$
and the interface transparency parameter $\gamma_b$. The first one is defined
as %%
\begin{equation}
\label{eq4}
\gamma=\frac{\rho_S \xi_S}{\rho_F \xi_F},
\end{equation}
the second one can be calculated from the critical thickness  of the
S layer $d^\mathrm{(crit)}_S$ which is defined as the thickness below which
there is no superconductivity in the S/F bilayer:
$T_c(d_S^\mathrm{(crit)})=0$.

In the limiting case $(\gamma/\gamma_b)(d_S/\xi_S)\ll 1$,  the
thickness $d_S^\mathrm{(crit)}$ can be calculated explicitly as
\cite{Fominov2002}
\begin{equation}
\label{eq5}
\frac{d_S^\mathrm{(crit)}}{\xi_S} = 2\gamma_E \frac{\gamma}{\gamma_b}.
\end{equation}
Here $\gamma_E \approx 1.78$ is the Euler constant. Our data yield
$d_S^\mathrm{(crit)}/\xi_S\simeq 0.8$ ($\gamma_b=1.95$) for the Fe/In system,
$d_S^\mathrm{(crit)}/\xi_S\simeq 1$ ($\gamma_b=2.7$) for the Fe/Cu/Pb system,
and $d_S^\mathrm{(crit)}/\xi_S\simeq 0.7$
($\gamma_b=0.37$) for the HA/Cu/Pb system. All obtained parameters are presented in Table~\ref{tc_tab}.
The larger value of $\gamma_b$ for the Fe/Cu/Pb system than for the Ha/Cu/Pb system makes of course sense.
Indeed, the difference between the HA system and the Fe system is seen in a difference of $\gamma_b$,
that helps to rationalize the use of a weaker F-layer.
\begin{table}
\caption{Parameters used for fitting of the theory to the
experimental  results (see, e.g., \cite{Leksin2015})}.
\begin{center}
\begin{tabular}{|c|c|c|c|} \hline

 & 1& 2& 3 \\ %\hline

 \cline{2-4}
\raisebox{1.5ex}[0cm][0cm]{Parameter }

 & Fe2/In & Fe2/Cu/Pb & HA/Cu/Pb \\ \hline
$\rho_S$, $\mu \Omega \cdot$cm &
0.2 & 1.47 & 1.47 \\ \hline
$l_S$, nm &
 300 & 17 & 17 \\ \hline
$D_S$, cm$^2$/s &
 1100 & 100 & 100 \\ \hline
$\xi_S$, nm &
170 & 41 & 41 \\ \hline
$\rho_F$, $\mu \Omega \cdot$cm &
10 & 10 & 130 \\ \hline
$l_F$, nm &
10 & 10 & 6.41 \\ \hline
$D_F$, cm$^2$/s &
 3.3 & 3.3 & 21.4 \\ \hline
$\xi_F$, nm &
7.5 & 7.5 & 14 \\ \hline
$\xi_h$, nm &
 0.5 & 0.3 & 1.25 \\ \hline
$\gamma$ &
0.45 & 0.78 & 0.03 \\ \hline
$\gamma_b$ &
1.95 & 2.7 & 0.37 \\ \hline
\end{tabular}
\label{tc_tab}
\end{center}
\end{table}

Fig.~\ref{HAdepend} summarizes the experimental $\Delta T_c(d_{\rm
F2 })$  dependences for SSV heterostructures with HA as the F2 layer
obtained in the present work and our previous results on the
Fe-based SSVs  \cite{Leksin2012,Leksin2015}. Solid lines in
Fig.~\ref{HAdepend} are theoretical results using the parameters
listed in Table~\ref{tc_tab}. The general feature of the SSVs with F2\,=\,Fe is
that the measured points at small thicknesses $d_{\rm F2 }$ lie much
lower than the theoretically expected positive maximum of $\Delta
T_c$ (direct SSV effect). One should note that the difference in the
theoretical maximum values of  $\Delta T_c$ for In- and  Pb-based
systems (inset in Fig.~\ref{HAdepend}) is caused by the different values of the
superconducting transition temperature of the single S layer
($T_{c}$\,=\,3.4\,K for In and $T_{c}$\,=\,7.18\,K for Pb).
Obviously in both cases, to reach the expected maximum it would be
necessary to further decrease the thickness of the Fe2 layer. It
should be emphasized that the thickness $d_{\rm F2 }$ is one of the
crucial parameters for the functionality of the spin valve. It
determines the number of the Cooper pairs which experience the
influence of the exchange fields of \emph{both} F layers in the
heterostructure.
In general, to get the maximal magnitude of the
spin-valve effect $\Delta T_c$, the thickness $d_{F2}$ of the F2
layer proximate to the S layer should be of the order or smaller than the
penetration depth of the Cooper pairs into the F2 layer
$\xi_h=\sqrt{\hbar D_F/h}$. Here $h$ is the exchange
splitting of the conduction band of a ferromagnet. The thinner F2 layer is,
the more Cooper pairs can reach the contact region between the two
ferromagnetic layers where at certain thicknesses of the F1 and F2
layers the compensation effect of the exchange fields can take place in the AP case. For
the previously studied Fe-based systems $h$ was of the order of
$\sim 1$\,eV and $\xi_h$ amounted to $0.6 - 0.8$\,nm
\cite{Leksin2012,Leksin2015}. According to theory \cite{Fominov2010}
this means that the maximum of $\Delta T_c$ should occur in the
interval between 0.3 and 0.4\,nm [Fig.~\ref{HAdepend}(inset)]. With
the available experimental setup it is practically impossible to
grow a continuous iron film of such small thickness.

It is well known \cite{Kontos1,Kontos2} that in dilute alloys, e.g. in PdNi alloys
with 10 \% of Ni, $\xi_h$ is of the order of 5\,nm, which is
an order of magnitude larger as compared  to pure ferromagnetic
elements such as Fe, Ni, or Co.

As our present experimental results demonstrate, the use  of the
Heusler alloy for the growth of the F2 layer is very beneficial. It
greatly relaxes the stringent condition on the minimum thickness of
the F2 layer. Indeed, according to the previous analysis of the
Pb/Cu/Co$_2$Cr$_{1-x}$Fe$_x$Al trilayers  the HA film grown at the
substrate temperature of $T_{\rm sub}= 300$\,K can be classified as
a weak ferromagnet with a relatively small exchange field $h^{\rm
HA}\sim 0.2$\,eV \cite{Kamashev2017}. As can be seen in
Fig.~\ref{HAdepend},  such reduction of $h$ shifts the peak of the
theoretical $\Delta T_c(d_{\rm F2 })$ dependence for F2\,=\,HA to
larger thicknesses of the order of 1\,nm which can be easily
reached experimentally. Under these conditions the measured maximum
magnitude of $\Delta T_c$ is two times larger compared to the
best previous result on the Fe-based SSVs (Fig.~\ref{HAdepend}). In
fact, it almost reaches the theoretically predicted value
suggesting that further optimization of the properties of the F2
layer is unlikely to significantly increase the SSV effect. In this
respect it would be very interesting to explore theoretically and
experimentally the option of optimization of the F1 layer in the SSV
AFM/F1/N1/F2/N2/S heterostructure.

Recently, Singh \textit{et al.} \cite{Aarts2015} reported a huge SSV effect
for the S/F1/N/F2 structure made of amorphous MoGe, Ni, Cu and CrO$_2$ as S, F1, N and F2, respectively.
This structure demonstrated $\Delta T_c $ for $\sim 1$ K when changing the relative orientation of
magnetizations of two F layers. The reason for such surprisingly strong  SSV effect
remains unclear \cite{Ouassou2017}. Gu {\it, etc.} \cite{Gu1, Gu2} reported $\Delta T_c \sim 400 $ mK for the three-layer Ho/Nb/Ho films.

Finally, a discrepancy between the theoretical curves  and
experimental data at larger $d_{F2}$ thicknesses in the regime of
the inverse (negative) SSV effect found for all the above discussed
systems (Fig.~\ref{HAdepend}) needs to be commented.
%We suppose that the main  reason for the deeper minimum observed at the negative
%values of $\Delta T_c$ as compared to theory could be a finite
%transparency of the F1/Cu/F2 interface for the Cooper pairs which is
%not taken into account in theory \cite{Fominov2010}. Further factors
In this respect, we note that the assumptions of theory \cite{Fominov2010}
do not fully comply with the properties of our samples.
While the assumption of F layers being weak ferromagnets (exchange energy much
smaller than the Fermi energy) is satisfied for the Heusler alloy, iron is closer
to the limit of strong ferromagnets (exchange energy starts to be comparable with the Fermi energy).
Accurate theoretical description of ferromagnets with large exchange splitting requires taking into account
different densities of states in different spin subbands and modified boundary conditions at SF interfaces
\cite{Cottet2009,Eschrig2015}.
At the same time, the major inconsistency between theory and experiment in our case is probably related to
the assumption of the dirty limit (mean free path much smaller than the coherence length).
In our samples,
these assumptions are close the border of applicability or even not satisfied (depending
on the specific material). The ferromagnets turn out to be strong enough so that the condition
$l_F \ll \xi_h$ is not satisfied at all.
Therefore, we cannot expect theory \cite{Fominov2010} to describe quantitative details of our results.
Still, we observe that the theory captures main qualitative features of the experiment.
%could be the use of a simplified model of the superconducting and
%ferromagnetic metals as well as the fact that the starting points of
%the theory do not strictly comply with the properties of our
%samples. Indeed, the F layers were assumed to be weak ferromagnets.
%The dirty limit for the F layer considered by
%theory is also not realized owing to a small value of $\xi_F$.
%
%It is necessary to note that the theoretical fits in the original papers
%\cite{Leksin2012,Leksin2015} are performed using the parameter of the theory
%$W$ which as it turned out later describes experimental points better.
%
%We also note that usually the  F1 layer is considered  in
%theory as semi-infinite \cite{Leksin2015} because  $\Delta T_c$ has
%only a minor maximum at $d_{F1}\sim\xi_h/2$.

\section{Conclusion}

In summary, we have experimentally demonstrated that using  for the
F2 layer in the CoO$_x$/F1/Cu/F2/Cu/Pb heterostructure a specially
prepared thin film of the Heusler alloy Co$_2$Cr$_{1-x}$Fe$_x$Al
with a small degree of the spin polarization of the conduction band
significantly increases the magnitude of the superconducting spin
valve effect $\Delta T_c$ as compared to similar systems with the F2
layer made of the strong ferromagnet Fe. It follows from our
theoretical analysis that the experimentally achieved value is close
to the maximum predicted by theory. The use of the Heusler alloy
did not increase this maximum value beyond the theoretical result but enables
to reach experimentally  the maximum possible $\Delta T_c$ at a larger,
technically realizable thickness of the F2 layer, in a full agreement with theory.
It seems unlikely that further optimization of the material of the
F2 layer would yield substantially larger values of  $\Delta T_c$.
A potentially interesting alternative would be to optimize the
parameters of the F1 layer which is tempting to explore in future.

\begin{acknowledgements}
A.K.\ gratefully acknowledges the Leibniz - DAAD
Fellowship.
Ya.F. was partially supported by the RF Ministry of Education and Science
(grant No.\ 14Y.26.31.0007), by the RFBR (grant No.\ 17-52-50080),
and by the Basic research program of HSE.
\end{acknowledgements}

\bibliography{HA_SpinValve}

\end{document}